 \let\accentvec\vec
\documentclass[twocolumn]{svjour3}          
 \let\spvec\vec

\usepackage{mathptmx}
\usepackage{graphicx}
\let\vec\accentvec
\usepackage{amsmath}
\let\vec\spvec
\usepackage{eufrak}
\usepackage{hyperref}
\hypersetup{
	linkcolor=blue,
	citecolor=blue,
	filecolor=blue,
	urlcolor=blue,
	colorlinks=true
}
\usepackage[export]{overpic}
\usepackage[usenames,dvipsnames]{xcolor}
\usepackage{xr}

\usepackage{color}
\usepackage{xcolor}
\usepackage{soul} 
\soulregister\cite7
\soulregister\ref7
\usepackage{comment}
\renewcommand{\v}[1]{\vec{#1}} 
\newcommand{\m}[1]{#1}

\newcommand{\plane}[2]{$#1$\nobreakdash-$#2$~plane}

\begin{document}\sloppy

\title{Theory of AC quantum transport with fully electrodynamic coupling\thanks{This work was supported by the National Science Foundation under
CAREER ECCS-1351871.}
}

\author{Timothy M. Philip \and Matthew J. Gilbert}


\institute{T. M. Philip \at
              Department of Electrical and Computer Engineering, University of Illinois at Urbana-Champaign, Urbana, IL 61801, USA \\
              \email{tphilip3@illinois.edu}           
           \and
           M. J. Gilbert \at
           \email{matthewg@illinois.edu}
}

\date{Received: \today / Accepted: date}

\maketitle

\begin{abstract}
    With the continued scaling of microelectronic devices along with the growing demand of high-speed wireless telecommunications technologies, there is increasing need for high-frequency device modeling techniques that accurately capture the quantum mechanical nature of charge transport in nanoscale devices along with the dynamic fields that are generated. In an effort to fill this gap, we develop a simulation methodology that self-consistently couples AC non-equilibrium Green functions (NEGF) with the full solution of Maxwell's equations in the frequency domain. We apply this technique to simulate radiation from a quantum-confined, quarter-wave, monopole antenna where the length $L$ is equal to one quarter of the wavelength, $\lambda_0$. Classically, such an antenna would have a narrower, more directed radiation pattern compared to one with $L \ll \lambda_0$, but we find that a quantum quarter-wave antenna has no directivity gain compared to the classical solution. We observe that the quantized wave function within the antenna significantly alter the charge and current density distribution along the length of the wire, which in turn modifies the far-field radiation pattern from the antenna. These results show that high-frequency radiation from quantum systems can be markedly different from classical expectations. Our method, therefore, will enable accurate modeling of the next generation of high-speed nanoscale electronic devices.
\end{abstract}


\section{Introduction}


The non-equilibrium Green function (NEGF) methodology has found great utility in modeling DC electron transport in a wide variety of quantum systems such as quantum dots~\cite{Klimeck2007,Huang2009,Balzer2009}, resonant tunneling diodes~\cite{Lake1992,Do2006}, and metal-oxide-semiconductor (MOS) transistors~\cite{Luisier2006,Fiori2007,Koswatta2007}. The success of the method derives not only from its ability to accurately model quantum coherent transport but also from its systematic framework within which to introduce incoherence and interactions~\cite{Lake1997,Datta2000,Anantram2008a,Pourfath2014}. The standard, DC formulation of the NEGF method, however, fails to capture transient features of transport and electrodynamic coupling that are important in phenomena such as cross-talk in high-frequency circuits~\cite{Grondin1999}. Although, such full-wave coupling has been treated semi-classically in the context of the time-domain Boltzmann transport and the drift-diffusion equations~\cite{Witzig1999,Grondin1999,Sirbu2005,Willis2010,Willis2011}, fully dynamic electromagnetic coupling to quantum transport has not yet been achieved.

The time-domain NEGF formulation can be used to understand dynamic phenomena and has successfully been applied to elucidate phenomena ranging from the AC conductance in graphene~\cite{Perfetto2010} to the transient properties of quantum interferometers~\cite{Gaury2014a,Tu2014,Tu2016}. Transient NEGF calculations, however, suffer from the memory effect, whereby the entire history of the Green function must be stored in order to accurately evolve the Green function~\cite{Myohanen2008,PuigvonFriesen2010}. Although some approximations, such as the wide bandwidth limit, can greatly reduce the storage requirements, time-dependent calculations are still demanding for time scales longer than a few picoseconds since a sub-femtosecond time step is required for numerical stability and the computation time scales at least linearly with the total number of time steps. Therefore, to accurately model device responses at sub-terahertz operating frequencies, the DC NEGF formulation has been extended to simulate AC steady state transport, thereby providing a platform for studying quantum transport at any frequency~\cite{Wei2009,Kienle2010,Shevtsov2013,Zhang2013}.

Although AC NEGF provides a simple extension of DC NEGF to understand high-frequency transport, care must be taken to accurately account for displacement current and thereby satisfy gauge invariance and charge conservation~\cite{Wei2009,Zhang2013}. The AC NEGF methodology on its own does not satisfy either of these conditions, and so they must be enforced through additional measures. Fortunately, this problem can be alleviated by self-consistently solving the transport equations with the Coulomb potential at the Hartree level of approximation~\cite{Wei2009,Zhang2013}. To date, however, electromagnetic interactions with AC NEGF transport have been assumed to be quasi-static, where a solution of Poisson's equation is sufficient to capture the electrostatics of the problem. The quasi-static approximation fails, however, in situations that involve dynamic electromagnetic fields, such as electromagnetic induction or field radiation, which are features that are vital to characterize the properties of high-frequency devices~\cite{Larsson2007}.

In this work, we  present a simulation technique that self-consistently solves the AC NEGF equations with the full solution of Maxwell's equations in the frequency domain to consider the influence of both static and dynamic fields in the presence of fully quantum mechanical electron transport in three dimensions.  In Sect.~\ref{sec:method}, we introduce details of the simulation methodology, including the tight-binding form of Hamiltonian and the governing equations of the method for quantum transport, which are used to calculate charge and current density within the system.  We then present a finite-difference frequency-domain (FDFD) method that, instead of solving for the electric and magnetic fields, solves directly for the scalar and vector electromagnetic potentials on a Yee cell.  By solving for the potential formulation of Maxwell's equations rather than the electric and magnetic fields that standard treatments obtain, we can directly couple the output potentials into the Hamiltonian. The output charge and current densities from the AC NEGF are coupled with the FDFD potentials to obtain a self-consistent solution. In Sec.~\ref{sec:results}, we apply this technique to model the electrodynamic radiation emitted by a quantum monopole antenna, that is an antenna that possesses quantum-confined  states. We begin with an overview of the classical operation of a monopole antenna, with emphasis on the expected radiation patterns at different operating frequencies. We then compare these classical expectations to the the AC NEGF/FDFD simulations of a quantum monopole. The quantization within the quantum wire significantly alters the charge and current density distribution, which ultimately changes the macroscopic radiation pattern. Despite driving the antenna at the quarter-wave frequency, we find that the quantum monopole antenna fails to achieve the directivity gain associated with a classical quarter-wave antenna. In Sec.~\ref{sec:conclusion}, we summarize and discuss the implication of our results.

\section{Simulation Methodology}\label{sec:method}

\subsection{Tight Binding Hamiltonian}

Our transport formulation models the quantum system under consideration with a nearest-neighbor real-space, tight-binding Hamiltonian of the form
\begin{equation}
  \mathcal H(\v r) = \sum_{\v{r}} \left[ \psi_{\v{r}}^\dag H_0\psi_{\v{r}} + \sum_{\v\delta}\left(\psi_{\v{r}}^\dag H_{\v\delta}\psi_{\v{r} + \v\delta} + \text{H.c.}\right) \right], \label{eq:TBHam}
\end{equation}
where $\psi_{\v{r}}$ is the electron annihilation operator at position $\v r$, $\psi^\dag_{\v{r}}$ is the electron annihilation operator at position $\v r$, $\v\delta$ are the distances between nearest neighbor atoms in the lattice, $H_0$ is the on-site term, and $H_{\v\delta}$ is the nearest-neighbor hopping term. For the case of the cubic lattice we consider in this work, $\v\delta$ has the form $\v\delta = (\pm a \,\v{\hat x}, \pm a \,\v{\hat y}, \pm a \,\v{\hat z})$, where $a$ is the lattice constant.

\subsection{AC NEGF} 

The governing equations for the AC NEGF derive from the time-domain Keldysh equation~\cite{Kadanoff1962,Pourfath2014}. The formal expression for the full, two-time Green function for the total system including the leads is given by the Dyson equation~\cite{Kadanoff1962,Stefanucci2013}
\begin{equation}
    \begin{split}
        G^\gamma(t_1, t_2) = &g_0^\gamma(t_1, t_2) +  \\
        \int dt_3 \int dt_4 \, &\left[ \,g_0^\gamma(t_1,  t_3)  \Sigma^\gamma(t_3, t_4) \delta(t_3-t_4) G^\gamma(t_4, t_2) , \right.\\
        &\,\, + \left. g_0^\gamma(t_1,  t_3)  U(t_3)\delta(t_3-t_4)  G^\gamma(t_4, t_2) \right],
    \end{split} \label{eq:td_GF_Full}
\end{equation}
where $G^\gamma(t_1,t_2)$ is the fully-dressed retarded or advanced Green function ($\gamma = \{r,a\}$), $g_0^\gamma(t, t')$ is the Green function for the isolated (i.e. uncontacted) device Hamiltonian including any time-independent potentials, $\Sigma^\gamma(t, t')$ is the self-energy that describes the time-independent and time-dependent influence of the leads, and $U(t)$ is the time-dependent potential energy.  Without loss of generality, Eq.~\eqref{eq:td_GF_Full} can be Fourier transformed to the energy domain by using the two-time Fourier transform defined as~\cite{Kienle2010}
\begin{equation}
    H(E_1,E_2) = \int \frac{dt_1}{2\pi} \int \frac{dt_2}{2\pi} \, e^{iE_1t_1/\hbar} e^{iE_2t_2/\hbar} h(t_1,t_2),
\end{equation}
where $h(t_1, t_2)$ is any generic time-domain function of times $t_1$ and $t_2$, and $H(E_1, E_2)$ is the Fourier transform of $h$ at energies $E_1$ and $E_2$. Since the AC bias implies that the system is no longer time-independent, we cannot utilize a single time/energy Fourier transform based on the difference $t_1-t_2$ as is traditionally considered in DC NEGF. 

The resultant energy-domain form of Eq.~\ref{eq:td_GF_Full} is written
\begin{equation}
    \begin{split}
        G^\gamma(E_1, E_2) &= G_0^\gamma(E_1) \delta( E_1 - E_2 ) + \\
                    &\int d E_3 \,G_0^\gamma(E_1) \Sigma^\gamma(E_1-E_3) G^\gamma(E_3, E_2) \\
                    &\int d E_3 \,G_0^\gamma(E_1) U(E_1-E_3) G^\gamma(E_3, E_2),
    \end{split}\label{eq:Ed_GF_Full}
\end{equation}
where $G_0^\gamma(E)$ is the DC contribution to the Green function given by
\begin{equation}
  G^r_0(E) = \left[ (E + i\eta) I - \mathcal H - U_0 - \Sigma_0^r(E) \right]^{-1}. \label{eq:Gr0}
\end{equation}
Here, $I$ is the identity matrix, $\eta$ is an infinitesimal positive number that pushes the poles of the Green function into the complex plane, allowing for integration along the real energy axis~\cite{Anantram2008a}, $U_0$ is the DC potential energy profile, and $\Sigma_0^r(E)$ is the DC contact self-energy that integrates out the influence of the semi-infinite leads. 
In general, the contact-self-energy is non-analytic, but decimation techniques can be used to efficiently obtain self-energies for semi-infinite leads~\cite{Sancho2000a,Sancho2000}.

To obtain the system response at frequency $\omega$, we evaluate the Green function in Eq.~\eqref{eq:Ed_GF_Full} at $E_1 = E_+ = E+ \hbar\omega$ and $E_2 = E$. For a AC bias of the form $V(t) = V_{AC} \cos \omega t$ applied to the leads, we simplify the expression for the Green function to
\begin{equation}
  G^r(E) = G^r_0(E) + g^r_\omega(E), \label{eq:Gdef}
\end{equation}
where $g^r_\omega(E)$ is first-order response of the system to the AC bias. Similarly, the total self-energy can be expressed as a function of DC component and a small-signal AC component:
\begin{equation}
    \Sigma^\gamma(E) = \Sigma^\gamma_0(E) + \sigma^\gamma_\omega(E) \quad (\gamma = r, <), \label{eq:Sigmadef}
\end{equation}
where $\sigma^\gamma_{\omega}(E)$ is the AC self-energy due to a the small-signal AC bias. Previous work has shown that this AC self-energy can be perturbatively expanded in powers of $e V_{AC}/\hbar\omega$, where $e$ is the electron charge, which provides a systematic framework to evaluate the system's response~\cite{Shevtsov2013} Using this result, the AC self-energy to first-order in the bias amplitude is given simply as
\begin{equation}
  \sigma^\gamma_{\omega}(E) = \frac{eV_{AC}}{\hbar\omega} [\Sigma_0^\gamma(E) - \Sigma_0^\gamma(E_+)] \quad (\gamma = r, <). \label{eq:sigma}
\end{equation}
Taking advantage of this simplified form, the AC retarded Green function, $g^r_\omega(E)$, can be determined by substituting Eqs.~\eqref{eq:Gdef}-\eqref{eq:sigma} into Eq.~\eqref{eq:Ed_GF_Full} and only retaining terms that are linear in the bias amplitude, $V_{AC}$. The resulting expression is a product of DC Green functions at energies $E$ and $E_+ = E+\hbar\omega$~\cite{Wei2009}: 
\begin{equation}
  \begin{split}
  g^r_\omega(E) = G^r_0&(E_+) \left[ U_\omega + \sigma^r_{\omega}( E)  \right]  G^r_0(E),
  \end{split} \label{eq:grw}
\end{equation}
where $U_\omega$ is the AC potential energy profile. The AC retarded Green function is sufficient to calculate density of states and, in the wide-band limit when the contact self-energy is assumed to be independent of energy, terminal currents, but the AC lesser Green function is necessary to calculate spatially-resolved particle and current density.

\begin{figure}[t]
\begin{center}
  \includegraphics[width=\columnwidth]{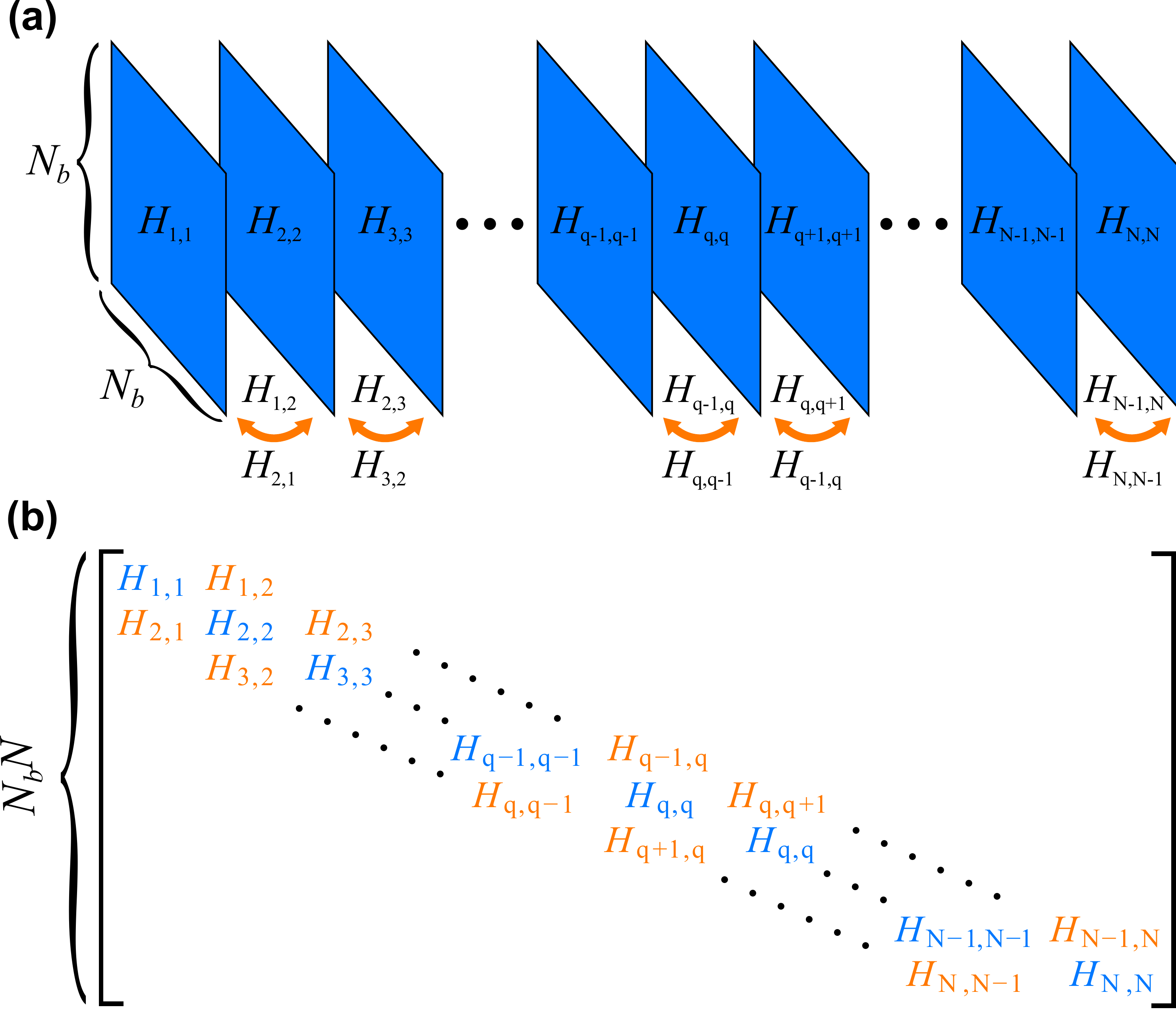}
\end{center}

\caption[]{ (a) The device region can be considered diagonal matrix slices $H_{q,q}$ connected to the nearest neighbor lattice $H_{q+1,q+1}$  through off diagonal matrix blocks $H_{q,q+1}$ and $H_{q+1,q}$ (b) The tight-binding Hamiltonian matrix of this layered hopping structure is a block tridiagonal matrix. The AC recursive Green algorithm exploits this matrix structure to efficiently calculate the AC Green functions.\label{fig:RGF_Hamiltonian}} 
\end{figure}

The AC lesser Green function gives one direct access to charge flow within the system by convolving the density of states information contained in the AC retarded Green function with the occupancy of the leads that is stored in the lesser self-energy, $\Sigma^<(E)$. In the energy-domain, this convolution can be simply written by the Keldysh equation $G^<(E) = G^r(E) \Sigma^<(E) G^r(E)^\dag$~\cite{Wei2009}. After Eqs.~\eqref{eq:Gdef} and~\eqref{eq:Sigmadef} are substituted into this relation, the expression for the AC lesser Green function to first-order in the AC bias amplitude is given as
\begin{equation}
    \begin{split}
        g_\omega^<(E) = &\, G_0^r(E_+)\Sigma_0^<(E_+) g_\omega^r(E)^\dag \\
                        &+ G_0^r(E_+)\sigma_\omega^<(E) G_0^r(E)^\dag \\
                        &+ g_\omega^r(E)\Sigma_0^<(E) G_0^r(E)^\dag.
    \end{split} \label{eq:gnw}
\end{equation}
Again, we see that the AC linear response of the system is a product of DC Green functions at energies $E$ and $E_+$.

Once the AC retarded Green function and AC lesser Green function are computed, observables can be calculated in a fashion similar to DC NEGF~\cite{Anantram2008a,Pourfath2014}.  The spatially-resolved local density of states (LDOS) is only dependent on the AC retarded Green function and is calculated by 
\begin{equation}
  \text{LDOS}(\v r) = \frac{i}{2\pi} \left( g^r_{\omega;\v r, \v r}(E) - g^r_{\omega;\v r, \v r}(E)^\dag\right).
\end{equation}
 The frequency-dependent electron density $n_\omega(\v r)$ is given as
\begin{equation}
  n_\omega(\v r) = -i\int\frac{ dE}{2\pi}\, g^<_{\omega; \v r, \v r} (E) . \label{eq:electron_density}
\end{equation}
While the electron density is important to understand charge dynamics, the AC particle current density must be determined to self-consistently calculate the dynamic magnetic field within the system and  is given by
\begin{equation}
    \begin{split}
        J_{\alpha,\omega}(\v r) = &- \frac{e}{\hbar} \int \frac{dE}{2\pi}  \left[ H_{\v r + a\v{\hat \alpha} , \v r} g^<_{\omega; \v r, \v r + a\v{\hat \alpha}}(E)\right. \\
         &\left.- g^<_{\omega; \v r + a\v{\hat \alpha}, \v r}(E) H_{\v r, \v r + a\v{\hat \alpha} }\right]  \quad (\alpha = x,y,z). \label{eq:current_density}
    \end{split}
\end{equation}
Additionally, the AC terminal currents can be computed as \cite{Wei2009}
\begin{equation}
    \begin{split}
        I_{\beta,\omega} = -\frac{e}{\hbar} &\int \frac{dE}{2\pi} \\
        \text{Tr}\left[\vphantom{G^<_0}\right.   &g^<_{\omega}(E) \Sigma^r_{0\beta}(E)^\dag - \Sigma^r_{0\beta}(E_+)  g^<_{\omega}(E)  \\
        &g^r_{\omega}(E) \Sigma^<_{0\beta}(E) - \Sigma^<_{0\beta}(E_+) g^r_{\omega}(E)^\dag \\
        &G^r_{0}(E_+) \sigma^<_{\omega\beta}(E) - \sigma^<_{\omega\beta}(E) G^r_{0}(E)^\dag\\
        &G^<_{0}(E_+) \sigma^r_{\omega\beta}(E)^\dag - \sigma^r_{\omega\beta}(E)  G^<_{0}(E)    \left.\vphantom{G^<_0}\right].
    \end{split} \label{eq:current}
\end{equation}
where $\beta = L,R$ for the left and right contact currents.

\subsection{AC Recursive Green Function Algorithm}\label{app:recursive_algorithm}

Although the AC NEGF equations described in the previous section can be directly solved to simulate electron transport, it is a computationally intensive task since two matrix inversions are required to obtain  $G^r_0(E)$ and $G^r_0(E_+)$ at each step of the energy integration necessary to calculate the particle density and current density in Eqs.~\eqref{eq:electron_density} and \eqref{eq:current_density}. For a matrix of side length $N_\text{tot}$, the number of operations needed to invert it scales as $O(N_\text{tot}^3)$~\cite{Anantram2008a}. Since the Hamiltonian side length is a product of the number of lattice sites and the number of orbitals, the AC NEGF algorithm becomes prohibitively expensive for large systems due to the computational burden of inverting large matrices. Calculating particle and current density using Eqs.~\eqref{eq:electron_density} and \eqref{eq:current_density}, however, does not require all of the matrix elements of the entire Green function. In fact, only the main diagonal blocks and the first off-diagonal blocks of the lesser Green function are necessary to calculate these observables. With this in mind, we develop a AC recursive Green function (RGF) algorithm for the AC NEGF equations that efficiently solves for only the main diagonal and first off-diagonal of the AC lesser Green function while avoiding full matrix inversions by exploiting the sparse nature of the Hamiltonian matrix. Our AC RGF algorithm extends the well-known DC RGF algorithm that is commonly used in modern DC NEGF transport simulations to the AC NEGF Eqs.~\eqref{eq:grw} and \eqref{eq:gnw}~\cite{Lake1997,Anantram2008a}.

Since our algorithm requires some parts of the DC RGF algorithm, we begin by reviewing the RGF algorithm for the DC NEGF method before extending it to the AC case. Equation~\eqref{eq:Gr0} for the DC retarded Green function, $\m G^r$, can be rewritten as
\begin{equation}
  \mathcal A \m G^r = \m I, \label{eq:AGI}
\end{equation}
where $\mathcal A = ((E+i\eta)\m I - \mathcal H - U_0 - \m \Sigma^r )$ is the inverse of the Green function, $\mathcal H$ is the  matrix representation of the Hamiltonian, $U_0$ is the DC potential energy profile,  and $\m \Sigma^r$ is the self-energy that captures the influence of external leads and and scattering mechanisms. In order to avoid inverting the entire $\mathcal A$ matrix to obtain the Green function, we must consider the structure of the Hamiltonian. The RGF algorithm is most efficiently applied to layered systems, where the Hamiltonian matrix parallel to the transport direction consists of $N_b$ orbitals/lattice sites that can be considered a single layer or slice. Each $N_b \times N_b$ slice of the Hamiltonian is connected only connected to the next slice in the transport direction though an off-diagonal $N_b \times N_b$ block. Aside from the these main  diagonal block and first off-diagonal blocks, the rest of the Hamiltonian matrix $\mathcal H$ is zero. Figure~\ref{fig:RGF_Hamiltonian} schematically depicts the layered system and the resulting Hamiltonian matrix. We adopt a matrix notation where $H_{q,s}$ refers to a $N_b \times N_b$ block of the Hamiltonian, $\mathcal H$, for block row index $q$ and block column index $s$:
\begin{equation}
	H_{q,s} = \mathcal H_{ (q-1)N_b+1: qN_b,  (s-1)N_b+1:s N_b }
\end{equation}
For example, when $N_b = 10$, $H_{3,4}$ refers to the matrix block comprising of rows 21 to 30 and columns 31 to 40.  In Fig.~\ref{fig:RGF_Hamiltonian}a, we depict the each layer of $N_b$ orbitals and lattice sites as blue slices labeled by $H_{q,q}$, where $q$ runs from 1 to $N$ total number of slices. Coupling between the layers in the transport direction is indicated by the arrows between the slices labeled $H_{q+1,q}$ and $H_{q+1,q}$. Figure~\ref{fig:RGF_Hamiltonian}b depicts the matrix representation of this layered structure as a sparse, block tridiagonal matrix with side length $N_\text{tot} = N_b N$, where all nonzero blocks are labeled by their block indices. The $\mathcal A$ matrix in Eq.~\eqref{eq:AGI} for these layered structures is, therefore, also sparse and has the same block tridiagonal structure as the Hamiltonian. The DC RGF algorithm exploits the sparsity and block tridiagonal form of this matrix to calculate the relevant blocks of the Green function using $N$ inversions of $N_b \times N_b$ and a number of computationally trivial matrix multiplications instead of inverting a single matrix of side length $N_b N$. This reduces the computational complexity of calculating the Green function from $O(N_b^3N^3)$  to $O(N_b^3N)$ and provides significant speedup in obtaining the Green function for systems that have many layers~\cite{Lake1997,Anantram2008a}.

\subsubsection{Retarded Green Function}

The RGF algorithm begins by inverting the first $N_b \times N_b$ block of the $\mathcal A$ matrix, which is the first block of the left-connected Green function, $\mathfrak g^{r,L}$. The name of this Green function derives from the fact that it only contains information about the left-lead through the contact self-energy that is part of the first block of $\mathcal A$. To obtain the elements of the full Green function, we must build the rest of $\mathfrak g^{r,L}$ until it is connected to the right contact self-energy. The blocks of a given Green function will be subscripted with the same notation as the $\mathcal A$ matrix. The first block of the left-connected Green function is given simply as 
\begin{equation}
  \mathfrak g^{r,L}_{1,1} =  \mathcal A_{1,1}^{-1}.  \label{eq:grL1}
\end{equation}
Subsequent diagonal blocks of $\mathfrak g^{r,L}$ are connected back to the first block recursively through the following relationship:
\begin{equation}
  \mathfrak g^{r,L}_{q,q} =  \left( \mathcal A_{q,q} - \mathcal A_{q,q-1} \mathfrak g^{r,L}_{q-1,q-1} \mathcal A_{q-1,q} \right)^{-1}.  \label{eq:grLq}
\end{equation}
Thus, $\mathfrak g^{r,L}$ is constructed by started on the top left of the $\mathcal A$ matrix and iterating over $q$ until $q = N$, where the right contact is located. Since this last block, $\mathfrak g^{r,L}_{N,N}$, includes the influence of the contact self energy at both $q=1$ and $q=N$, it is exactly the last diagonal block of the fully-connected Green function $\m G^r_{N,N}$.

We also build an analogous right-connected Green function that begins at $q = N$ and is connected only to the right contact though the contact self-energy that is present in $\mathcal A_{N,N}$. The $q = N$ block is given as 
\begin{equation}
	\mathfrak g^{r,R}_{N,N} =  \mathcal A_{N,N}^{-1}.  \label{eq:grR1} \\
\end{equation}
The blocks of the right-connected Green function are found by the recursive relationship
\begin{equation}
	\mathfrak g^{r,R}_{q,q} =  \left( \mathcal A_{q,q} - \mathcal A_{q,q+1} \mathfrak g^{r,R}_{q+1,q+1} \mathcal A_{q+1,q} \right)^{-1}.  \label{eq:grRq}
\end{equation}
Therefore, the blocks of $\mathfrak g^{r,R}$ are computed by iterating backwards from $q = N$ to $q = 1$. Again, we note that similar to the last block of $\mathfrak g^{r,L}$ being the last block of the full Green function, the $q = 1$ block of right-connected Green function, $\mathfrak g^{r,R}_{1,1}$, is exactly $\m G^r_{1,1}$. In the DC NEGF formulation, one typically only needs to compute one of either $\mathfrak g^{r,L}$ or $\mathfrak g^{r,R}$ to obtain the desired elements of the full Green function, but for the AC extension of the recursive algorithm described in the following sections, we will need both to obtain the leftmost and rightmost columns of the full Green function. Although these columns can be calculated using just $\mathfrak g^{r,L}$ and matrix multiplication, the large number of off-diagonal components of the Green that must be computed makes the approach computationally inefficient. It is, therefore, more efficient to calculate both $\mathfrak g^{r,L}$ and $\mathfrak g^{r,R}$ to obtain the desired columns.

Once the left-connected and right-connected Green functions have been constructed, any off-diagonal block of the full $G^r$ matrix can be computed using multiplicative relationships. Elements above the main diagonal of the Green function are recursively related by
\begin{equation}
	\begin{split}
    \left. \m G^{r}_{q,s} \right|_{q < s} &= - \mathfrak g^{r,L}_{q,q} \mathcal A_{q,q+1} G^{r}_{q+1,s} \\
    &= - G^r_{s,q-1}  \mathcal A_{q-1,q}  \mathfrak g^{r,R}_{q,q},
  \end{split} \label{eq:GrOffUpper} 
\end{equation}
and elements below the main diagonal follow the relationship
\begin{equation}
	\begin{split}
    \left. \m G^{r}_{q,s} \right|_{q > s} &=  -  G^{r}_{s,q+1}  \mathcal A_{q+1,q} \mathfrak g^{r,L}_{q,q} \\
      &=  - \mathfrak g^{r,R}_{q,q} \mathcal A_{q,q-1} G^{r}_{q-1,s}. 
  \end{split} \label{eq:GrOffLower}
\end{equation}


The AC RGF implementation requires the left and right columns of $\m G^r$, so we use Eqs.~\eqref{eq:GrOffUpper}-\eqref{eq:GrOffLower} to obtain them by multiplying away from the main diagonal blocks:
\begin{align}
   \m G^{r\text{(rgt)}}_{q,N} &= - \mathfrak g^{r,L}_{q,q} \mathcal A_{q,q+1} G^{r\text{(rgt)}}_{q+1,N},  \label{eq:GrRgt} \\
   \m G^{r\text{(lft)}}_{q,1} &= - \mathfrak g^{r,R}_{q,q} \mathcal A_{q,q-1} G^{r\text{(lft)}}_{q-1,1}.   \label{eq:GrLft} 
\end{align}
For clarity, we add the superscript (rgt) to the elements of the rightmost column of the Green function and (lft) to the elements of the leftmost column. Once the left and right columns of the DC Green function are obtained, the AC Green function can be computed.

\subsubsection{RGF Formulation for AC Retarded Green Function}

For the AC RGF algorithm described below, we assume coherent transport such that AC self-energies, $\sigma^{r}_\omega$ and $\sigma^{<}_\omega$, are only nonzero in the first ($q = 1$) and last ($q = N$) diagonal sub-blocks due to the presence of contacts. Under such an assumption, we show in the proceeding section that the lesser AC Green function, $g^{<}_{\omega; q,q}$, blocks can be calculated using simple matrix multiplication using just the left and right block columns of $G^{r}_0$, $G^{r}_+$, and $g^{r}_\omega$. 

The AC NEGF method involves calculating the product of Green functions at two different energies, previously denoted as $\m G^r_0(E)$ and  $\m G_0^r(E+\hbar\omega)$. As such, there will be a number left-connected and right-connected Green functions associated with each full matrix. To simplify notation, we will denote functions associated with $\m G^r_0(E)$ with a subscript 0 and those with $\m G_0^r(E+\hbar\omega)$ with subscript +. For example, $\mathfrak g^{r,L}_0$ refers to the left-connected Green function of $\m G^r_0(E)  = G^{r}_0 $, while $\mathfrak g^{r,L}_+$ refers to that of $\m G^r_0(E+\hbar\omega)  = G^{r}_+$. The AC Green function will continue to be subscripted with $\omega$, so, $\mathfrak g^{r,L}_\omega$ refers to left-connected Green function of the AC retarded Green function, $g^{r}_\omega$.

We can rewrite Eq.~\eqref{eq:grw} in this abbreviated notation as
\begin{align}
  \m g^{r}_\omega &= \m G^{r}_+( \sigma^{r}_{\omega} + U_\omega )\m G^{r}_0, \\
  \mathcal A_+ \m g^{r}_\omega &=  ( \sigma^{r}_{\omega} + U_\omega )  \m G^{r}_0, \label{eq:grAC}
\end{align}
where $\mathcal A_+ = ((E+\hbar\omega + i\eta)\m I - \m H - \m \Sigma(E+\hbar\omega) )$. Equation~\eqref{eq:grAC} resembles the equation of the DC lesser Green function $\m G^{<}_0$~\cite{Datta2000}, and so, we develop an RGF algorithm for $g^{r}_\omega$, similar to the recursive algorithm used to calculate the DC lesser Green function~\cite{Anantram2008a}. To obtain $g^{r}_\omega$, we first use the recursive algorithm from the previous section to obtain the left- and right-connected Green functions for both $G^{r}_0$ and $G^{r}_+$ in addition to the left and right columns of their full matrices.

To calculate the relevant blocks of the AC retarded Green function, we first must obtain the diagonal blocks of the left-connected AC retarded Green function, $\mathfrak g^{r,L}_{\omega;q,q}$. The first, $q=1$, block of $\mathfrak g^{r,L}_{\omega;q,q}$ can be trivially calculated from the left-connected Green functions $\mathfrak g^{r,L}_{0}$ and $\mathfrak g^{r,L}_{+}$:
\begin{equation}
  \mathfrak g^{r,L}_{\omega;1,1} = \mathfrak g^{r,L}_{+;1,1} (\m U_{\omega;1,1} + \m \sigma^{r}_{\omega;1,1}) \mathfrak g^{r,L}_{0;1,1}. \label{eq:grwL1}
\end{equation}
The subsequent diagonal blocks of of the $\mathfrak g^{r,L}_\omega$ can be computed through the multiplicative, recursive relationship
\begin{equation}
  \begin{split}
  \mathfrak g^{r,L}_{\omega;q,q} = &\mathfrak g^{r,L}_{+;q,q} \left[ \m U_{\omega;q,q} + \m \sigma^{r}_{\omega;q,q} \right.  \\
    &\quad\left. + \mathcal A_{+;q,q-1} \mathfrak g^{r,L}_{\omega;q-1,q-1} \mathcal A_{+;q-1,q}  \right] \mathfrak g^{r,L}_{0;q,q}. 
  \end{split}\label{eq:grwLq}
\end{equation}
In general, for any pair of block indices $q$ and $s$ where $q\neq s$, $\mathcal A_{+;q,s} = \mathcal A_{q,s}$, so no additional matrix elements must be computed to evaluate Eq.~\eqref{eq:grwLq}. We simply retain the subscripted $+$ to keep consistent notation with Eq.~\eqref{eq:grAC}. We also compute the right-connected AC retarded Green function, $\mathfrak g^{r,R}_\omega$, via analogous expressions:
\begin{align}
  \mathfrak g^{r,R}_{\omega;N,N} &= \mathfrak g^{r,R}_{+;N,N} [\m U_{\omega;N,N} + \m \sigma^{r}_{\omega;N,N}]\mathfrak g^{r,R}_{0;N,N}  \label{eq:grwR1} \\
  \begin{split}
  \mathfrak g^{r,R}_{\omega;q,q} &= \mathfrak g^{r,R}_{+;q,q} [ \m U_{\omega;q,q}  + \m \sigma^{r}_{\omega;q,q}  \\
  &\quad +\mathcal A_{+;q,q+1} \mathfrak g^{r,R}_{\omega;q+1,q+1} \mathcal A_{+;q+1,q}  ]\, \mathfrak g^{r,R}_{0;q,q}.
  \end{split} \label{eq:grwRq}
\end{align}

To calculate the lesser AC Green function, we need the leftmost and rightmost columns of $\m g^{r}_\omega(E)$. We can obtain them using the minimum number of matrix multiplication through the following relations:
\begin{align}
  \begin{split}
  \m g^{r\text{(rgt)}}_{\omega; q,N} = &
    - \mathfrak g^{r,L}_{\omega; q,q} \mathcal A_{ q, q+1}^{+} \m G^{r\text{(rgt)}}_{0; q+1,N}  \\
    &\qquad- \mathfrak g^{r,L}_{+; q,q}     \mathcal A_{ q, q+1}^{+}           \m g^{r\text{(rgt)}}_{\omega; q+1,N},   
  \end{split}\label{eq:GrwRgt} \\
  \begin{split}
  \m g^{r\text{(lft)}}_{\omega; q,1} = &
    - \mathfrak g^{r,R}_{\omega; q,q} \mathcal A_{ q, q-1}^{+}  \m G^{r\text{(lft)}}_{0; q-1,1} \\
    &\qquad- \mathfrak g^{r,R}_{+; q,q}     \mathcal A_{ q, q-1}^{+}           \m g^{r\text{(lft)}}_{\omega; q-1,1}. 
  \end{split}\label{eq:GrwLft}
\end{align}
Again, we add the superscript (rgt) to the elements of the right column of the Green function and (lft) to the elements of the left column. Using this methodology, we simplify the AC Green function computation from two inversions of $N_b N \times N_b N$ matrices to $2N$ inversions of $N_b \times N_b$ matrices along with a handful of computationally trivial matrix multiplications. This reduces the computational complexity from $O(N_b^3 N^3)$ to $O(N_b^3 N)$, which can greatly increase computational speed.

\subsubsection{AC Lesser Green Function}

Once the leftmost and rightmost columns of the AC retarded Green function are obtained using Eq.~\eqref{eq:GrwRgt}-\eqref{eq:GrwLft}, we can calculate the AC lesser Green function, which is then used calculate relevant observables. In the abbreviated notation we have established, Eq.~\eqref{eq:gnw} for the AC lesser Green function is written as
\begin{align}
    g^{<}_\omega &= G^{r}_+\Sigma^{<}_+ g^{r\dag}_\omega + G^{r}_+\sigma^{<}_{\omega} G^{r\dag}_0  + g^{r}_\omega\Sigma^{<}_0 G^{r\dag}_0, \\
  				&= g^{<1}_\omega + g^{<2}_\omega + g^{<3}_\omega.
\end{align}
In the second line, we label each term in the first line with a superscripted number. It is more convenient to calculate each of these terms individually and then add them together to obtain the total AC lesser Green function.

If the two-terminal transport is coherent, the self-energies $\Sigma^<_{0/+}/\sigma^<_\omega$ are zero except for the $q=1$ block diagonal and the $q = N$ block diagonal due to the presence of contact self-energies. With such simplification, the three terms of $g^{<}_\omega$ can be calculated using block matrix multiplication involving the leftmost and rightmost columns of $G^r_{0/+}/g^r_\omega$.  In the block notation of the RGF algorithm, the three terms of $g^{<}_\omega$ are calculated as
\begin{align}
  \begin{split}
  g^{<1}_{\omega;q,q} =  
    \m G^{r\text{(lft)}}_{+;q,1}  &\Sigma^{<}_{+;1,1} \left( \m g^{r\text{(lft)}}_{\omega;q,1} \right)^\dag \\
    +&\, \m G^{r\text{(rgt)}}_{+;q,N} \Sigma^{<}_{+;N,N} \left(\m g^{r\text{(rgt)}}_{\omega;q,N} \right)^\dag,  \label{eq:Gnw1qq}
  \end{split}\\
  \begin{split}
  g^{<2}_{\omega;q,q} = 
    \m G^{r\text{(lft)}}_{+;q,1}  &\sigma^{<}_{\omega;1,1} \left( \m G^{r\text{(lft)}}_{0;q,1} \right)^\dag \\
    +&\, \m G^{r\text{(rgt)}}_{+;q,N} \sigma^{<}_{\omega;N,N} \left(\m G^{r\text{(rgt)}}_{0;q,N} \right)^\dag, 
  \end{split}\\
  \begin{split}
  g^{<3}_{\omega;q,q} = 
    \m g^{r\text{(lft)}}_{\omega;q,1}  &\Sigma^{<}_{0;1,1} \left( \m G^{r\text{(lft)}}_{0;q,1} \right)^\dag \\
    +&\, \m g^{r\text{(rgt)}}_{\omega;q,N} \Sigma^{<}_{0;N,N} \left(\m G^{r\text{(rgt)}}_{0;q,N} \right)^\dag,
  \end{split}
\end{align}
Which when added together to obtain $g^{<}_{\omega;q,q}$, allows us to compute the particle density using Eq.~\eqref{eq:electron_density}. To compute current density using Eq.~\eqref{eq:current_density}, we need the off-diagonal elements $g^{<}_{\omega;q,q+1}$ and $g^{<}_{\omega;q+1,q}$. The off-diagonal components of $g^{<}_{\omega}$ are computed in a similar fashion to the diagonal elements:
\begin{align}{}
  \begin{split}
  g^{<1}_{\omega;q+1,q} =  
    \m G^{r\text{(lft)}}_{+;q+1,1}  &\Sigma^{<}_{+;1,1} \left( \m g^{r\text{(lft)}}_{\omega;q,1} \right)^\dag \\
    +&\, \m G^{r\text{(rgt)}}_{+;q+1,N} \Sigma^{<}_{+;N,N} \!\left(\m g^{r\text{(rgt)}}_{\omega;q,N} \right)^\dag\!, 
  \end{split}\\
  \begin{split}
  g^{<2}_{\omega;q+1,q} = 
    \m G^{r\text{(lft)}}_{+;q+1,1}  &\sigma^{<}_{\omega;1,1} \left( \m G^{r\text{(lft)}}_{0;q,1} \right)^\dag \\
    +&\, \m G^{r\text{(rgt)}}_{+;q+1,N} \sigma^{<}_{\omega;N,N} \left(\m G^{r\text{(rgt)}}_{0;q,N} \right)^\dag\!, 
  \end{split}\\
  \begin{split}
  g^{<3}_{\omega;q+1,q} = 
    \m g^{r\text{(lft)}}_{\omega;q+1,1}  &\Sigma^{<}_{0;1,1} \left( \m G^{r\text{(lft)}}_{0;q,1} \right)^\dag \\
    +&\, \m g^{r\text{(rgt)}}_{\omega;q+1,N} \Sigma^{<}_{0;N,N} \left(\m G^{r\text{(rgt)}}_{0;q,N} \right)^\dag,
    \end{split}
\end{align}
\begin{align}
  \begin{split}
  g^{<1}_{\omega;q,q+1} =  
    \m G^{r\text{(lft)}}_{+;q,1}  &\Sigma^{<}_{+;1,1} \left( \m g^{r\text{(lft)}}_{\omega;q+1,1} \right)^\dag \\
    +&\, \m G^{r\text{(rgt)}}_{+;q,N} \Sigma^{<}_{+;N,N} \left(\m g^{r\text{(rgt)}}_{\omega;q+1,N} \right)^\dag, 
  \end{split}\\
  \begin{split}
  g^{<2}_{\omega;q,q+1} = 
    \m G^{r\text{(lft)}}_{+;q,1}  &\sigma^{<}_{\omega;1,1} \left( \m G^{r\text{(lft)}}_{0;q+1,1} \right)^\dag \\
    +&\, \m G^{r\text{(rgt)}}_{+;q,N} \sigma^{<}_{\omega;N,N} \left(\m G^{r\text{(rgt)}}_{0;q+1,N} \right)^\dag,
  \end{split}\\
  \begin{split}
  g^{<3}_{\omega;q,q+1} = 
    \m g^{r\text{(lft)}}_{\omega;q,1}  &\Sigma^{<}_{0;1,1} \left( \m G^{r\text{(lft)}}_{0;q+1,1} \right)^\dag \\
    +&\, \m g^{r\text{(rgt)}}_{\omega;q,N} \Sigma^{<}_{0;N,N} \left(\m G^{r\text{(rgt)}}_{0;q+1,N} \right)^\dag, \label{eq:Gnw3qq1}
  \end{split}
\end{align}
If transport on incoherent or other blocks of the self-energies are nonzero, Eqs.~\eqref{eq:Gnw1qq}-\eqref{eq:Gnw3qq1} are no longer valid. Instead, $g^{<}_\omega$ must be constructed from left-connected and right-connected Green functions for each term $g^{<1/2/3}_\omega$ following the recursive framework used to calculate the elements of $g^{r}_\omega$, as presented in the previous section. Evaluating the above equations to obtain the elements of  $g^{<}_{\omega}$ only requires matrix multiplications, which are significantly less computationally demanding than matrix inversions. As such the computational complexity for this AC NEGF algorithm to obtain $g^{<}_{\omega}$ remains $O(N_b^3 N)$. 

\subsubsection{AC RGF Solution Procedure}

In summary, the RGF procedure to obtain the relevant blocks of the AC retarded and lesser Green functions within the limit of coherent, two-terminal transport can be calculated with the following steps:
\begin{enumerate}
	\item Evaluate the blocks of left-connected DC Green function using Eq.~\eqref{eq:grL1} at $E$, the energy of interest, to obtain $\mathfrak g^{r,L}_{0;1,1}$ and at $E_+= E + \hbar\omega$ to obtain $\mathfrak g^{r,L}_{+;1,1}$.
	\item Using the recursive relationship for the left-connected DC Green function  defined in Eq.~\eqref{eq:grLq}, construct the remaining blocks of $g^{r,L}_{0}$ and $g^{r,L}_{+}$ by iterating $q = 2,3, \dots N-1, N$.
	\item Evaluate the blocks of the right-connected DC Green function using Eq.~\eqref{eq:grR1} at $E$ to obtain $\mathfrak g^{r,R}_{0;N,N}$ and at $E^+$ to obtain $\mathfrak g^{r,R}_{+;N,N}$.
	\item To obtain the the remaining blocks of the right-connected DC Green functions $\mathfrak g^{r,R}_{0}$ and $\mathfrak g^{r,R}_{+}$, iterate over $q= N-1,N-2,\dots,2,1$ and evaluate Eq.~\eqref{eq:grLq}. This iterative loop can also be used to calculate the rightmost columns of both fully-connected DC Green functions $G^{r}_0$ and $G^{r}_+$ using Eq.~\eqref{eq:GrRgt}.
	\item Now that all the blocks of the left- and right-connected DC Green functions $\mathfrak g^{r,L}_{0}$, $\mathfrak g^{r,L}_{+}$, $\mathfrak g^{r,R}_{0}$ and $\mathfrak g^{r,R}_{+}$, have been obtained, the AC left- and right-connected Green functions, $\mathfrak g^{r,L}_{\omega}$ and  $\mathfrak g^{r,R}_{\omega}$, can be computed. Start by using Eq.~\eqref{eq:grwL1} to calculate the first block of the left-connected AC Green function, $\mathfrak g^{r,L}_{\omega;1,1}$.
  	\item Build the remainder of the left-connected AC Green function, $\mathfrak g^{r,L}_{\omega}$, by iterating Eq.~\eqref{eq:grwLq} over $q=2,3,\dots,N-1,N$. It is also convenient to calculate the left columns of the fully-connected DC Green functions $G^{r}_0$ and $G^{r}_+$ using Eq.~\eqref{eq:GrLft} on this iterative loop.
  	\item Next, begin calculating the blocks of the right-connected AC Green function $\mathfrak g^{r,R}_{\omega}$ by using Eq.~\eqref{eq:grwR1} to obtain $\mathfrak g^{r,R}_{\omega;1,1}$.
  	\item Construct the remainder of the right-connected AC Green function $g^{r,R}_{\omega}$ by iterating Eq.~\eqref{eq:grwRq} over $q = N-1, N-2, \dots, 2,1 $. The right column of the fully-connected AC Green function $g^{r}_\omega$ can also be calculated during the loop using Eq.~\eqref{eq:GrwRgt}.
  	\item Finally, one last iterative loop over $q=2,3,\dots,N-1,N$ is necessary to calculate the the leftmost column of the fully-connected AC Green function $g^{r,R}_{\omega}$ using Eq.~\eqref{eq:GrwRgt}.
  	\item Now that the leftmost and rightmost columns of DC Green function at $E$, $G^{r}_0$, the DC Green function at $E_+$, $G^{r}_+$, and the AC Green function, $g^{r}_\omega$, have been calculated, Eqs.~\eqref{eq:Gnw1qq}-\eqref{eq:Gnw3qq1} are used to obtain the main diagonal and first off-diagonal blocks of the AC lesser Green function, $g^{<}_\omega$.
\end{enumerate}

This procedure must be repeated over a discretized energy range to compute particle density using Eq.~\eqref{eq:electron_density} with the main diagonal of $g^{<}_\omega$ and the current density using Eq.~\eqref{eq:current_density} with the first off-diagonal of $g^{<}_\omega$.

\begin{figure}[t]
\begin{center}
  \includegraphics[width=.8\columnwidth]{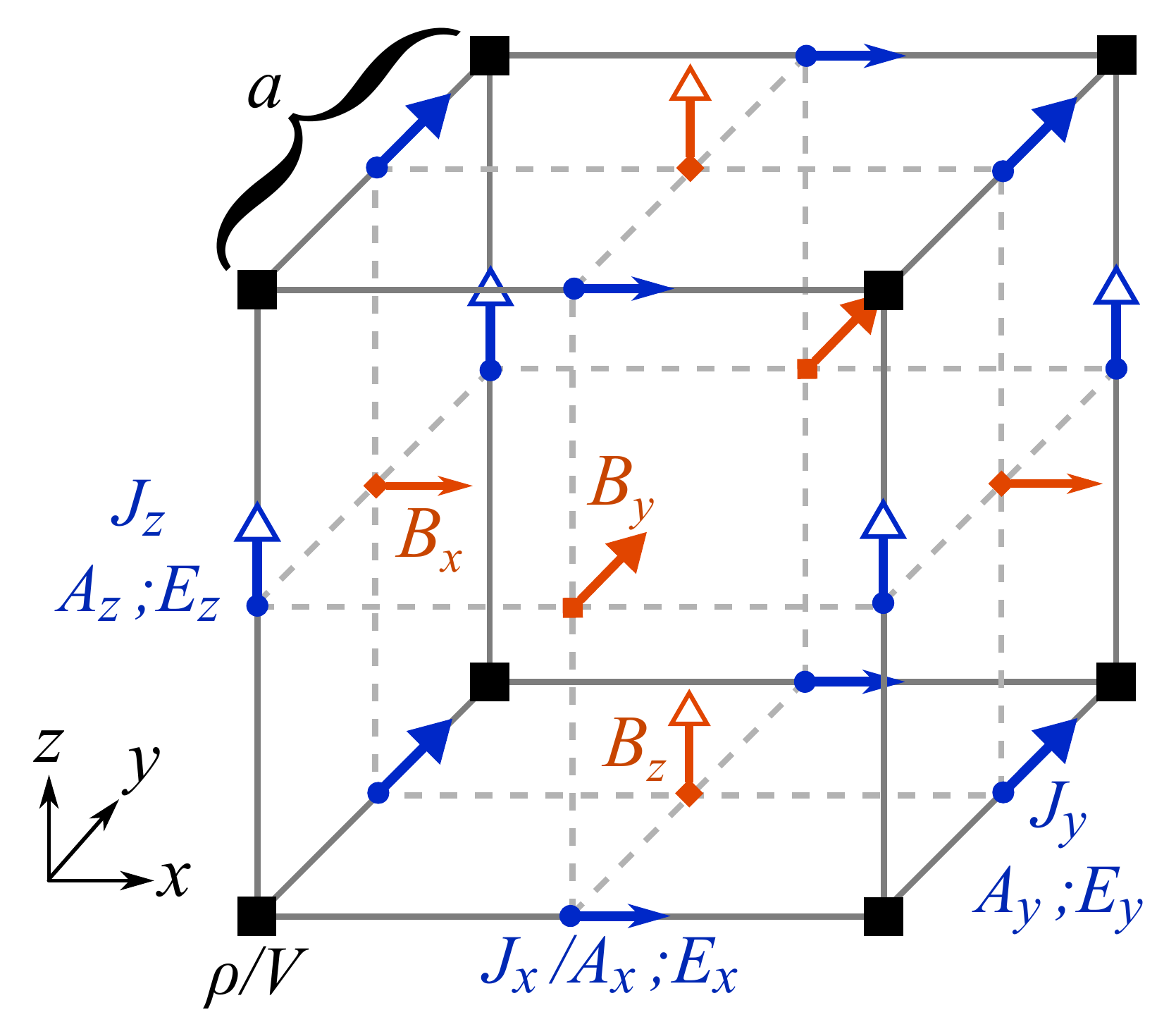}
\end{center}

\caption[]{The FDFD equations are solved on a cubic Yee cell with side length $a$, which matches the lattice constant of the tight-binding Hamiltonian. The tight-binding Hamiltonian lattice where the charge density, $\rho$, is located by the solid squares at the corners of the grid; the potential, $V$, is naturally defined at the same positions. The current density, $J_\alpha (\alpha = x, y, z)$, is located between lattice sites in the $\hat \alpha$ direction, so the same component of the vector potential $A_\alpha$ is also at these positions. The definition of electric field, $\v E = - \nabla \phi + i\omega \v A $, naturally co-locates $E_\alpha$ at the same sites. The definition of the magnetic flux density, $\v B = \nabla\times \v A$, places its components  at offset positions from the electric field.  \label{fig:yee_cell}}
\end{figure}

\subsection{Electrodynamics Coupling}

\subsubsection{Potential formulation of Maxwell's equations}

Charge transport through a system is strongly influenced by electric and magnetic fields, and thus must be accounted for when solving the AC NEGF equations. In DC NEGF, the self-consistent electrostatic potential is sufficient to capture the the effect of the electromagnetic environment. At frequencies when the ratio of the frequency to the speed of light, $\omega/c$, is not negligible, however, this electrostatic assumption is not adequate to account for the dynamic charge motion, so the full solution of Maxwell's equation must be solved to full understand the influence of electrodynamic coupling. 

Standard treatments for solving Maxwell's equations calculate the electric field, $\v E$, and magnetic field, $\v B$~\cite{Taflove2005}, but the quantum mechanical wave function is dependent, however, on the scalar potential $V$ and vector potential $\v A$~\cite{Chew2014}. We, therefore, reformulate the finite-difference frequency-domain (FDFD) method~\cite{Yee1966,Luebbers1990} to solve directly for the electromagnetic potentials. In the frequency-domain, Maxwell's equations in the Lorenz gauge, where $\nabla\cdot \v A = -\frac{i\omega}{c} V$, take the form
\begin{align}
  \left(\nabla^2 + \frac{\omega^2}{c^2} \right) V_\omega &= -\frac{\rho_\omega}{\varepsilon}, \label{eq:Vwaveeqn} \\
  \left(\nabla^2 + \frac{\omega^2}{c^2} \right) \v A_\omega &= -\mu \v J_\omega, \label{eq:Awaveeqn}
\end{align}
where $\omega$ is the frequency of interest, $c$ is the speed of light, $\varepsilon$ is the electric permittivity, and $\mu$ is the magnetic permeability. The charge density $\rho_\omega$ and current density $\v J_\omega$ are extracted from AC NEGF using Eqs.~\eqref{eq:electron_density} and \eqref{eq:current_density}. The electric field and magnetic flux density components, $E_\alpha$ and $B_\alpha$ $(\alpha = x,y,z)$, respectively, are numerically calculated from the potentials using the frequency domain relations:
\begin{align}
  \v E_\omega &= -\nabla V_\omega + i\omega  \v A_\omega, \label{eq:E}\\
  \v B_\omega &= \nabla \times \v A_\omega. \label{eq:B}
\end{align}

\subsubsection{Discretization}

Solving for the scalar and vector potential via Eqs.~\eqref{eq:Vwaveeqn} and \eqref{eq:Awaveeqn} on a discrete lattice requires care because the charge density and current density components from AC NEGF are not co-located on the same grid. The charge density is located at the lattice sites of the Hamiltonian, but the current density components, calculated from the particle flux transfered between lattice sites, is located between lattice sites. Therefore, Eqs.~\eqref{eq:Vwaveeqn} and \eqref{eq:Awaveeqn} must be solved on the staggered Yee cell illustrated in Fig.~\ref{fig:yee_cell}, where the scalar potential is defined at the same sites as the tight-binding lattice, while the components of the vector potential $A_\alpha\;(\alpha = x, y, z)$ are offset in position from the lattice to be co-located with the current density components $J_\alpha$. As a result, when Eqs.~\eqref{eq:E} and \eqref{eq:B} are evaluated using these staggered potentials, the electric and magnetic field naturally  are defined locations as would be expected in the standard finite-difference time-domain (FDTD) Yee cell discretization~\cite{Yee1966}.

Equations~\eqref{eq:Vwaveeqn} and \eqref{eq:Awaveeqn} both describe Helmholtz equations, which in rectangular coordinates can be generically written 
\begin{equation}
	 \left(\frac{\partial^2}{\partial x^2} + \frac{\partial^2}{\partial y^2} + \frac{\partial^2}{\partial z^2} + \frac{\omega^2}{c^2} \right) u(\v r) = - f(\v r). \label{eq:Helmholtz_eq}
\end{equation} 
To solve for the scalar potential, $f(\v r)$ is replaced with $\rho_\omega(\v r)/\varepsilon$ and $u(\v r)$ with $V_\omega(\v r)$. Similar replacements can be made for the different components of the vector potential. We discretize the Laplacian in Eq.~\eqref{eq:Helmholtz_eq} in rectangular coordinates using a second-order, central finite difference:
\begin{equation}
	\frac{\partial^2 u}{\partial x^2} \approx \frac{u_{i+1,j,k} - 2 u_{i,j,k} + u_{i-1,j,k}}{\Delta x^2},
\end{equation}
where we adopt the notation $u_{i,j,k} = u( i\Delta x, j\Delta y, k\Delta z)$. For uniform, cubic grid-spacing where $\Delta x = \Delta y = \Delta z$, the discrete form of the Helmholtz equation is written
\begin{equation}
	\begin{split}
	u_{i+1,j,k}  +  u_{i-1,j,k}  +  u_{i,j+1,k} +&\\
	 u_{i,j-1,k} +  u_{i,j,k+1} +  u_{i,j,k-1}+ &  \\
	\left(r_\omega^2 - 6 \right) u_{i,j,k} &= - \Delta{x}^2 f_{i,j,k},
	\end{split}\label{eq:disc_wave_eq_dx}
\end{equation}
where $r_\omega = \Delta x\omega/c$ and the $\Delta x$ is equal to the lattice constant, $a$, of the lattice Hamiltonian. To minimize discretization error, the condition $r_\omega < 1$ should be enforced~\cite{Wong2011}. This discretization forms a system of linear equations that can be solved by any number of linear algebra solvers.  

Appropriate boundary conditions must be applied at the edges of the simulation domain to accurately model the behavior of the fields away from the device region. To model a metallic boundary or ground place, we apply perfect electric conductor (PEC) boundary conditions that force the tangential components of the electric field to vanish. In the potential formulation of Maxwell's equation this boundary condition becomes~\cite{Chew2014}
\begin{align}
	\hat{\v n} \times \v A_\omega(\v r) &= 0 \\
			V_\omega(\v r) &= 0
\end{align}
where $\hat{\v n}$ is the unit normal to the surface of the boundary. The PEC boundary conditions perfectly reflect any incident electromagnetic waves, which accurately captures the behavior of highly conductive surfaces. When PEC boundary conditions are applied to all faces of the simulation domain, however, an electromagnetic cavity is created whereby waves with wavelengths commensurate with the domain side length are enhanced and others are attenuated. Such behavior must be avoided when modeling field radiation, as these cavity modes will significantly alter the radiation profile. As such, radiative boundary conditions that do not reflect incident waves must be applied for any boundary that is not explicitly metallic. To this end, stretched-coordinate, convolutional perfectly-matched-layers are attached to all non-metallic boundaries to absorb any radiating fields and inhibit the development of artificial cavity modes~\cite{Rickard2002,Shin2012}.

\subsubsection{Self-consistency with AC NEGF}

Once Eqs.~\eqref{eq:Vwaveeqn} and \eqref{eq:Awaveeqn} are solved using the charge and current density obtained from the AC NEGF calculations, the output scalar and vector potentials must be reinserted into the transport simulation. The scalar potential enters as an on-site potential which modifies the on-site term in the Hamiltonian described in Eq.~\ref{eq:TBHam}:
\begin{equation}
	H_0 \to H_0 - e V (\v r)
\end{equation}
The vector potential is coupled to the tight-binding Hamiltonian through a Peierl's phase on the off-diagonal hopping terms~\cite{Graf1995}: 
\begin{equation}
	H_{\v\delta} \to H_{\v\delta} \exp\left(-\frac{ie}{\hbar} \int_{\v r}^{\v r + \v \delta} \v A (\v r) \cdot d{\v \ell}\right).
\end{equation}
Based on the Yee cell discretization of the FDFD equations described in the previous section, the components of the vector potential are piece-wise constant between lattice sites. Therefore, the modified hopping amplitude is simplified to 
\begin{equation}
	H_\alpha \to H_\alpha \exp\left(\frac{-ie \Delta \alpha }{\hbar} A_\alpha(\v r + \Delta \alpha/2 \v{\hat \alpha} )\right),
\end{equation}
where $\alpha = \{x, y,z\}$. After these substitutions are made, the AC NEGF and FDFD equations must be solved again repeatedly until their solutions stabilize and self-consistency is attained. We mark self-consistency in our simulations when the change in the scalar potential between successive iterations is less than 1 $\mu$V. Reaching self-consistency with this iterative method can fail to converge and is often time-consuming, so we utilize the Anderson mixing scheme to stabilize and accelerate self-consistency~\cite{Bowler2000a}.

\begin{figure}[t]
  \begin{center}
    {\includegraphics[width=.9\columnwidth,clip,trim={0 .1in 2.2in .1in},page=1]{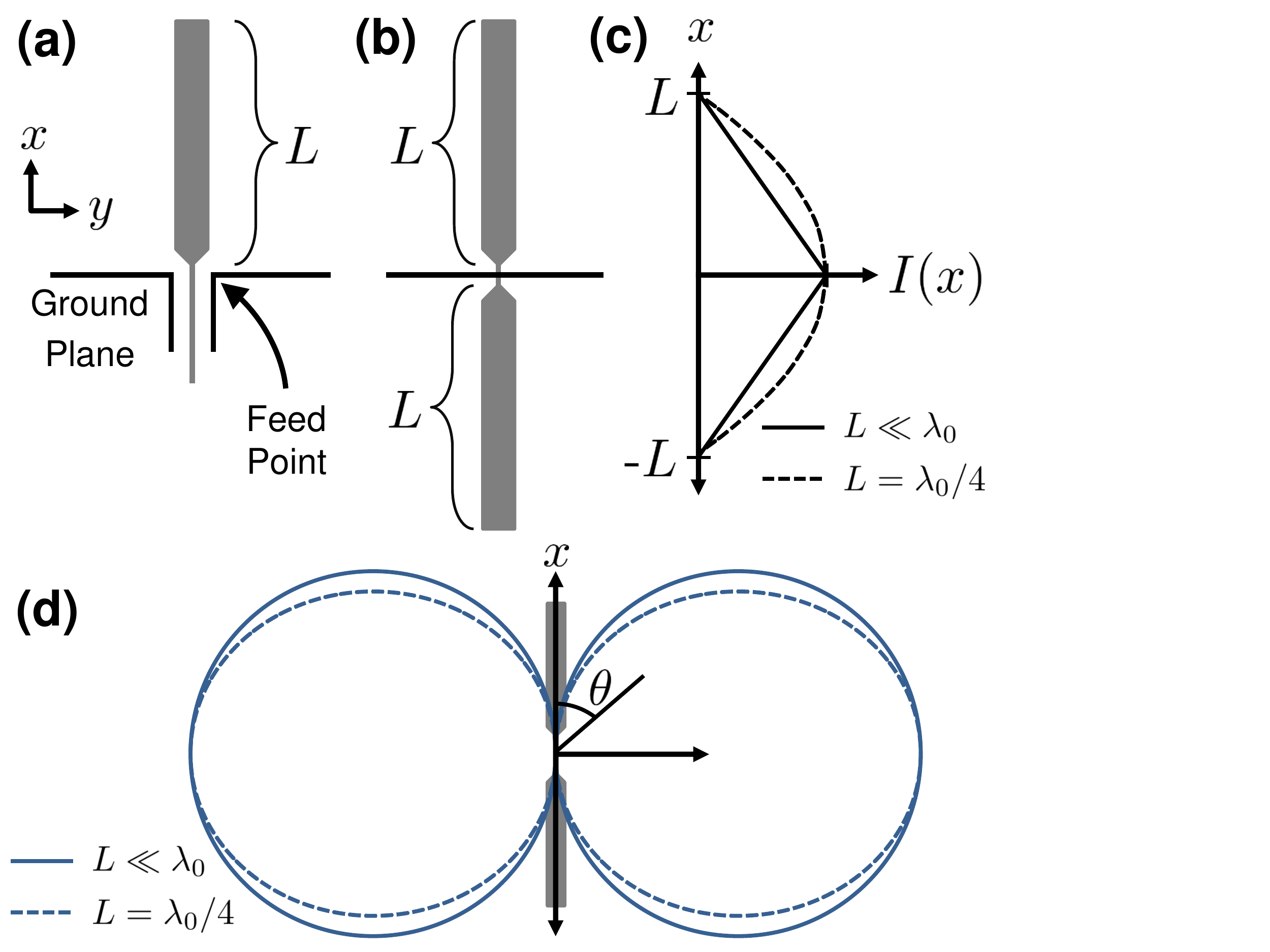}}
  \end{center}

  \caption[Schematic of the device design and the operation of a classical monopole antenna.]{Schematic of the device design and the operation of a classical monopole antenna. (a) A monopole antenna comprises of a current carrying wire of length $L$ that is extended above a conductive ground plane. (b) The image charge and currents in the ground plane results in radiation from the monopole antenna that mimics that from dipole antenna of length $2L$. (c) A short antenna is one in which the antenna length is much less than wavelength of operation ($L \ll \lambda_0$), which generates a current profile that linearly drops from the feed point to the end of the antenna. When the antenna length $L = \lambda_0/4$, the current distribution forms a sinusoidal, standing wave current profile. (d) The sinusoidal current profile of the quarter-wave monopole creates in a more directed classical radiation pattern than that of a short monopole. \label{fig:monopole_antenna}}
\end{figure}

\begin{figure}[t]
  \begin{center}
    {\includegraphics[width=.8\columnwidth,clip,trim={0 1.8in 3.8in 0},page=2]{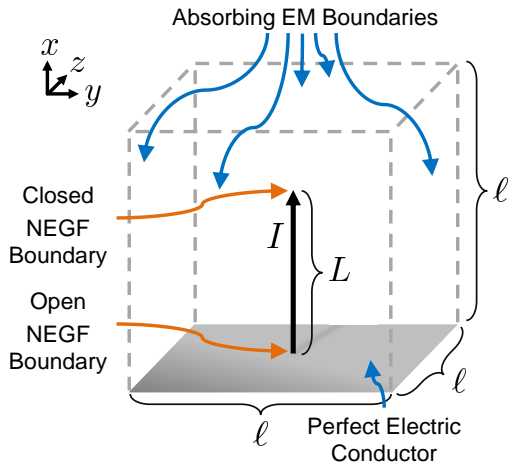}}
  \end{center}

  \caption[]{Schematic of simulated quantum monopole antenna. The AC NEGF technique is used to calculate transport for an end-fed 1D metal wire with $L=750$ $\mu$m. Transport is coupled self-consistently with the electrodynamic scalar and vector potentials in an FDFD domain with side length $\ell = 1.73$ mm. Perfect electrical conductor electromagnetic (EM) boundary conditions, indicated with the solid gray face, are applied to the bottom \plane{y}{z} while absorbing EM boundary conditions, indicated by the dashed edges, are applied on the other faces to inhibit the development of cavity modes. \label{fig:simulation_schematic}}
\end{figure}

\section{Monopole Radiation}\label{sec:results}

\subsection{Classical Case}\label{sec:classical}

To illustrate the efficacy and utility of this coupled AC NEGF/FDFD methodology, we numerically calculate the radiation emitted by a quarter-wave monopole antenna that possesses quantized energy states. Figure~\ref{fig:monopole_antenna}(a) illustrates the design of a monopole antenna. A wire of length $L$ is mounted on a conductive ground plane, which reflects the fields radiated from the antenna. The reflected fields from the ground plane can be modeled as image charges and currents below the monopole that result in radiation that is equivalent to a dipole antenna of length $2L$, as depicted in Fig.~\ref{fig:monopole_antenna}(b). 

Figure~\ref{fig:monopole_antenna}(c) illustrates the classically expected current density in an electrically short monopole antenna, that is where $L \ll \lambda_0$ and $\lambda_0$ is the operating wavelength, and a quarter-wave monopole antenna, where $L = \lambda_0/4$. In a short monopole, the current drops linearly from the feed point to the end of the antenna with $I_\text{short}(x) = i I_0 (L - x)$. When the antenna is extended to a length $L = \lambda_0/4$, however, the antenna operates at its resonant frequency and the current distribution forms a sinusoidal standing wave along the antenna with the form $I_{\lambda/4}(x) = i I_0 \cos \frac{2\pi x}{\lambda_0}$. 
Figure~\ref{fig:monopole_antenna}(d) shows the expected radiation pattern for both a short and quarter-wave monopole. We see that these altered current profiles along the length of the antenna result in different classical radiation patterns. The resultant far-field electric field for the short antenna has the form~\cite{HaytBuck2001}
\begin{equation}
  | E_\text{short} | =  \frac{ I_0 }{4 \varepsilon c r} \frac{ L}{\lambda_0}  \sin \theta,
\end{equation}
where $\theta$ is the angle measured from the $x$ axis. The angular dependence of radiation follows a simple $\sin\theta$ relationship that creates the circular lobes in the radiation pattern depicted in Fig.~\ref{fig:monopole_antenna}(d). When the antenna is driven at the quarter-wave frequency, the far-field electric field profile becomes
\begin{equation}
  | E_{\lambda/4} | =  \frac{ I_0 }{2\pi \varepsilon c  r} \frac{\cos \left(\pi/2\cos\theta \right) }{\sin \theta }.
\end{equation}
This more complicated angular dependence results in less power delivered at small angles and more at angles close to $\theta = \pi$, which is illustrated by the oval shaped lobes of the $\lambda_0/4$ radiation pattern in Fig.~\ref{fig:monopole_antenna}(d). Driving the antenna on resonance, therefore, creates a more directed radiation pattern than that of a short monopole, which can be used to more efficiently direct radiation at angles close to $\theta = \pi$.

\begin{figure}[t]
  \begin{center}
    {\includegraphics[width=.8\columnwidth,clip,trim={0 3.5in 4.8in 0}]{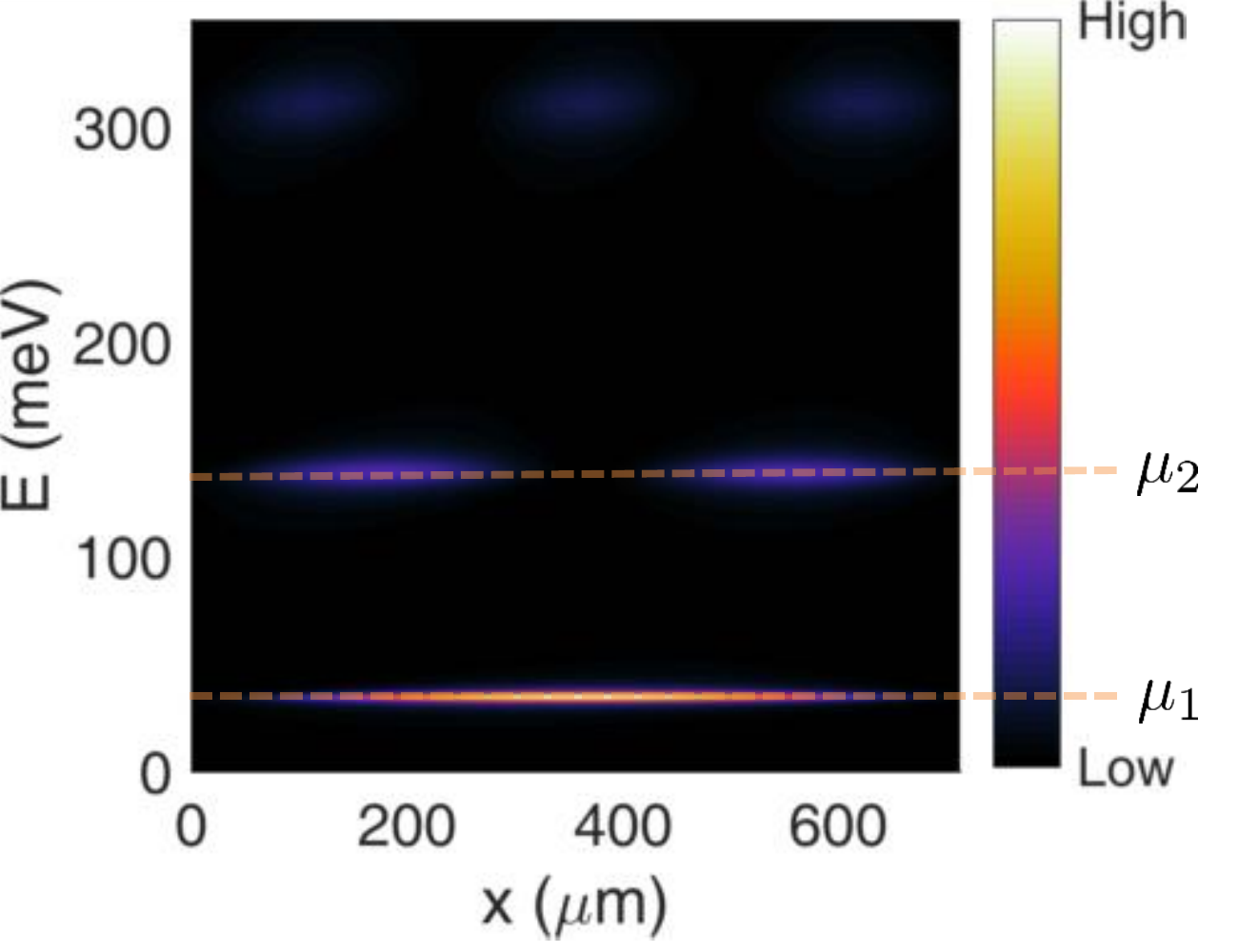}}
  \end{center}

  \caption[]{The calculated local density of states of the quantum wire antenna reveals quantized energy states at $\mu_1 = 34$ meV and $\mu_2 = 140$ meV. The quantum confined wave functions in this structure alters the current distribution within the antenna, which in turn modifies the macroscopic radiation pattern.   \label{fig:LDOS}}
\end{figure}

\begin{figure}[t]
  \begin{center}
    {\includegraphics[width=.8\columnwidth]{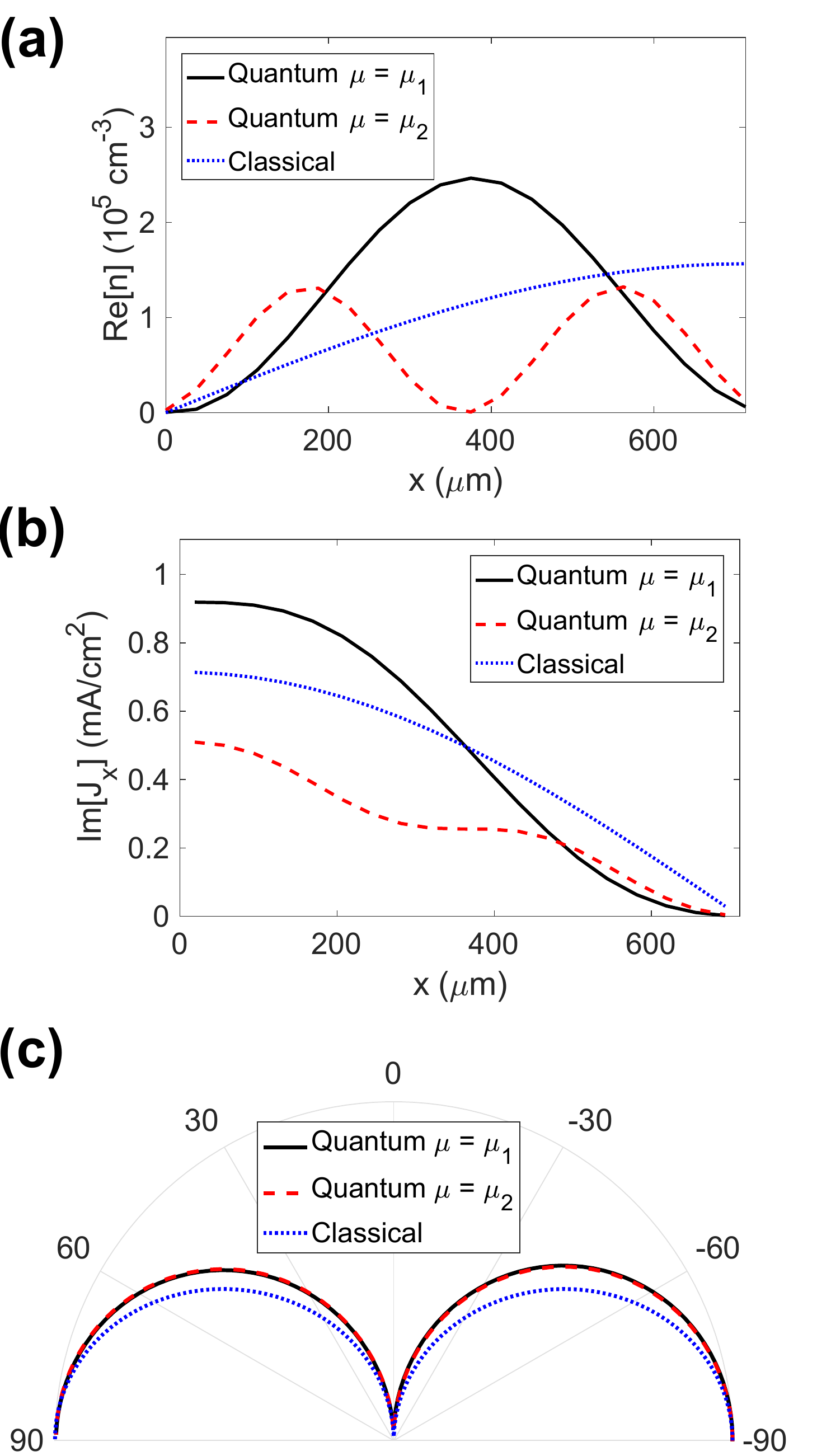}}
  \end{center}

  \caption[]{Simulation results of a quantum quarter-wave antenna through the first and second quantized energy level at the chemical potentials $\mu_1 = 34$ meV and $\mu_2 = 140$ meV, respectively. (a) The charge density distribution along the length of the antenna differs from the classical expectation due to the wave function of the quantized states. (b) The quantum confinement also alters the current density and does not create the expected sinusoidal profile. (c) The quantization effects on the charge and current density result in a modification to the macroscopic radiation pattern. \label{fig:results_E1}}
\end{figure}



\subsection{Quantum Case}\label{sec:quantum}

Having reviewed the expected radiation patterns from a classical short and quarter-wave monopole antenna, we now investigate the radiation behavior of a monopole antenna that possesses quantized energy states, which we will call a quantum monopole antenna. We use the self-consistent AC NEGF/FDFD technique described here to model the radiation of the quantum monopole antenna shown in Fig.~\ref{fig:simulation_schematic}. The antenna is modeled by a 20 lattice site one-dimensional metal Hamiltonian 
\begin{equation}
  \mathcal H(\v r) = \sum_{i} t_0\left[2  \psi_{x}^\dag \psi_{x} -  \left(\psi_{x}^\dag \psi_{x + a_0} + \text{H.c.}\right) \right], \label{eq:metal_Ham}
\end{equation}
with hopping amplitude $t_0 = 1.5$ eV and total length $L = 750$ $\mu$m~\cite{Datta2000}. Although the length of this antenna is relatively large and seemingly classical, the hopping parameter value we use creates quantized states within the wire. Since current is only injected at one end of the monopole antenna, only one end of the antenna has an open NEGF boundary condition, while the other end has a closed boundary condition.  The quantum antenna is placed in a cubic electromagnetics domain with side length $\ell = 1.73$ mm, with perfect electric conductor boundary conditions applied to the \plane{y}{z} to provide the ground plane needed for operation of the monopole antenna. Absorbing electromagnetic boundary conditions are applied to all other faces of the electrodynamics domain to allow for field radiation away from the antenna. With such a small electrodynamics simulation domain, the field surrounding the antenna is largely dominated by the non-radiative near field. To understand the far-field radiation pattern, we perform a near-to-far-field transformation on the self-consistent electrodynamic fields around the antenna~\cite{Petre1992}.  This technique allows us to understand the macroscopic radiative characteristics of the antenna without simulating a large electromagnetic domain outside the near field. We drive the antenna at the quarter-wave frequency of 100 GHz to understand the differences between the classical model of a monopole antenna using our quantum coherent AC NEGF/FDFD simulation methodology.

Figure~\ref{fig:LDOS} shows the computed LDOS of the quantum antenna. The energies of the states within the one-dimensional wire match those of an infinite square well within 5 meV, thereby demonstrating the quantum nature of transport within the antenna. We self-consistently calculate transport through the first and second quantized state by setting the chemical potential $\mu_1 = 34$ meV and $\mu_2 = 140$ meV, respectively. In Fig.~\ref{fig:results_E1}(a), we see that the AC quantum charge density differs significantly from the classically expected charge distribution. Rather than being maximized at the end of the antenna as is anticipated in the classical charge distribution, the wave function of the quantized electronic state maximizes the the charge distribution where the anti-nodes of the wave functions are seen in the LDOS. Because of the current-conserving nature of the AC NEGF/FDFD technique, the AC current density is intimately related to charge density via the continuity equation, and it too is altered from the expected classical arrangement. Despite being driven on the quarter-wave, resonant frequency, the current distribution no longer forms the standing wave along the length of the antenna. These non-ideal charge and current densities due to the spatial form of the quantum-confined wave function result in the distorted macroscopic radiation pattern in Fig.~\ref{fig:results_E1}(c). Therefore, instead of having the directivity of gain associated with a classical quarter-wave monopole, the quantum quarter-wave monopole has little to no gain at either chemical potential and radiates identically as a short antenna.


\section{Conclusion}\label{sec:conclusion}

In this work, we have detailed a novel quantum transport simulation methodology that goes beyond the quasi-static approximation by coupling AC NEGF with the full, dynamic solution of Maxwell's equations. This unique technique allows us to explore electrodynamic phenomena that cannot be understood via traditional DC or transient methods. We demonstrate the utility of this formulation by modeling the radiation from a quantum, quarter-wave, monopole antenna. We find that the quantum-confined wave function within the wire constrains the charge and current density profiles, which disallows the formation of the standing wave along the length of the wire despite being driven on resonance. Because of the altered current profile within the antenna, we observe that the macroscopic radiation pattern is altered by the quantization, resulting in no directivity gain compared to a short antenna. Our results illustrate that this new simulation methodology can uncover unexpected behavior in high-frequency quantum mechanical systems and will enable the characterization of the of next-generation, nanoscale, high-frequency devices, where quantum effects dominate transport phenomena. 


\section{Acknowledgments}
\noindent{}The authors acknowledge support from the NSF under CAREER Award ECCS-1351871. T.M.P. thanks M.J. Park and Y. Kim for helpful discussions.

\bibliographystyle{spphys}       
\bibliography{Antenna}

\end{document}